# Parylene Photonics: A Flexible, Broadband Optical Waveguide Platform with Integrated Micromirrors for Biointerfaces

Jay W. Reddy, Maya Lassiter, Maysamreza Chamanzar*
*Department of Electrical and Computer Engineering, Carnegie Mellon University, Pittsburgh, USA*

## ABSTRACT
Targeted light delivery into biological tissue is needed in applications such as optogenetic stimulation of the brain and in vivo functional or structural imaging of tissue. These applications require very compact, soft, and flexible implants that minimize damage to the tissue. Here, we demonstrate a novel implantable photonic platform based on a high-density, flexible array of ultracompact (30 μm x 5 μm), low-loss (3.2 dB/cm at λ = 680 nm, 4.1 dB/cm at λ = 633 nm, 4.9 dB/cm at λ = 532 nm, 6.1 dB/cm at λ = 450 nm) optical waveguides composed of biocompatible polymers Parylene C and polydimethylsiloxane (PDMS). This photonic platform features unique embedded input/output micromirrors that redirect light from the waveguides perpendicularly to the surface of the array for localized, patterned illumination in tissue. This architecture enables the design of a fully flexible, compact integrated photonic system for applications such as in vivo chronic optogenetic stimulation of brain activity.

## INTRODUCTION
From biological science to clinical practice, optical methods for imaging[1–3] and manipulation[4–9] of tissue are the gold standard for noninvasive interaction. However, the scattering and absorption of light in tissue pose fundamental limitations to the achievable resolution and depth of penetration. Depending on the type of tissue and wavelength of light, noninvasive optical microscopy techniques are typically limited to the superficial layers of tissue due to scattering and absorption of light, especially in the visible range of the optical spectrum[10]. Multiphoton techniques, which utilize simultaneous absorption of longer wavelength photons to activate visible-range optical agents, achieve deeper penetration into tissue due to reduced attenuation of near-infrared and infrared light. Even when using advanced multiphoton techniques, optical access is still limited to a couple of millimeters deep into the tissue[11]. In the context of brain imaging and manipulation, accessing deeper regions is required to study the neural mechanisms of disorders such as Parkinson's disease that involve malfunction of circuits in deep structures, namely, the basal ganglia nuclei. The issue of penetration depth is even more limiting for studying the large brains of nonhuman primates and humans. Implantable devices, which enable targeted light delivery deep into tissue, are therefore needed to advance scientific understanding of biological mechanisms and to aid clinical intervention.

Optical imaging and manipulation in free-roaming animal subjects require miniaturized technologies that are much smaller than traditional bulky microscopes. Recently demonstrated miniaturized microscopes and miniscopes, powered by electrical and fiber optic tethers, can be carried by a mouse during ambulatory experiments[12]. Moreover, compact optical implants, such as light-emitting diodes, optical fibers, and integrated photonic waveguides, have been used to deliver or collect photons deep within tissue to record and stimulate disparate regions simultaneously[13–20]. Unlike their external counterparts, these techniques require a physical device to be implanted into the tissue. Therefore, compact and flexible devices are highly desired to minimize damage to tissue while still benefiting from the power of optical techniques deep in tissue.

Among different biomedical applications, neurophotonics is an emerging field that demands minimally invasive and highly flexible optical implants for light delivery into the brain with high spatial resolution for optogenetic stimulation and functional imaging of brain activity. To study and understand the distributed and dynamic neural circuits in the brain, we need methods to monitor and manipulate neuronal activity at single-cell resolution over different areas of the brain during naturalistic behavior.

Brain tissue is especially vulnerable to damage from rigid implants. It has been shown that the performance of neural implants gradually degrades over time due to the foreign body response (FBR)[21]. This biological tissue response, which involves inflammation and astroglial scarring, is believed to be triggered, in part, by the mismatch between the mechanical properties of the implanted device and neural tissue[22]. The buildup of scar tissue around the implantation site degrades the recording signal-to-noise ratio and stimulation efficiency, limiting the lifetime of such implants. For electrical recording, flexible polymer devices have been shown to reduce damage to the brain tissue and thus enable longer-term neural recording[23]. An equivalent flexible optical platform is desired to enable optical interrogation of neural circuits.

Most existing integrated photonic waveguides are made of rigid semiconductor materials and dielectrics such as silicon, silicon nitride (SiN), and silicon dioxide [24]. These integrated photonic platforms are mainly designed for optical communications and are not necessarily optimized for implantable or wearable biophotonics. In addition to microfabricated integrated photonic devices, comparatively large single-channel light guides, including optical fibers and polymer silicone light guides[25,26], have been used for light delivery into tissue. Polymers such as SU-8 have also been incorporated into integrated photonic devices[15,27,28].

The overall stiffness of a device is determined by the geometry and Young's (elastic) modulus. The chronic tissue response is also a function of the shape of the probe cross-section. Typically, neural probe architectures are fabricated in long, high-aspect-ratio shapes that minimize the cross-sectional area to reduce acute tissue damage during implantation[29]. The probe mechanics in this shape are well characterized by the cantilever stiffness, which scales linearly with the Young's modulus of the material[21]. Compared to other commonly used materials for photonic waveguides, Parylene C and polydimethylsiloxane (PDMS) exhibit orders of magnitude lower Young's moduli, closer to the values of most biological tissues (Table 1), suggesting that this architecture will be less damaging to the surrounding tissue after implantation. Although polymer materials are still many orders of magnitude stiffer than the surrounding tissue, histology studies have shown that polymer probes cause a reduced foreign body response compared to more rigid silicon probes[30].

**Table 1:** Mechanical Properties of Tissues and Biomaterials

| Material | Young's Modulus, E (GPa) |
|---|---|
| Silicon[31] | 130 – 170 |
| Silicon Nitride[32] | 280 – 290 |
| Parylene C[33] | 1.5 – 4 |
| PDMS[34] | 1.32 – 2.97 x $10^{-3}$ |
| Brain Tissue[35] | (0.948 – 3.141) x $10^{-6}$ |
| Skin[36] | (6 – 222) x $10^{-6}$ |
| Muscle[36] | (2 – 12) x $10^{-6}$ |

Here, we demonstrate an integrated photonic platform to realize compact, biocompatible, and fully flexible polymer-based optical waveguide arrays that can deliver light efficiently into tissue in a minimally invasive way. Our architecture, Parylene photonics, is realized entirely in a flexible, biocompatible material platform composed of Parylene C and PDMS polymers. PDMS is optically transparent in the visible range and is resistant to degradation from prolonged exposure to a biological environment[37,38]. Both polymers are used in FDA-approved medical implants[39,40] and are also widely used in research as well as clinical applications[41]. With a proven history of biosafety in humans, Parylene C and PDMS form a compelling photonic platform for biointerface development, with translational potential for medical applications.

## DEVICE DESIGN AND ARCHITECTURE

### Compact Parylene photonic waveguides confine and guide light

Our flexible photonic platform, Parylene photonics, utilizes a dense array of waveguides in the shape of an implantable probe to deliver light from external light sources deep into tissue (Fig. 1a). Light is coupled to each waveguide from light sources located at the backend of the probe, which remains outside of the tissue. These light sources can be either integrated laser diodes or a fiber tether connected to an external laser source. The probe has a long flexible shank that is implanted into the tissue to deliver light at the target depth. Here, we have demonstrated 5 cm long waveguides. To minimize damage to the surrounding tissue, the shank is designed to be very thin (7 μm total thickness) and narrow (60 μm waveguide pitch). The total shank width is determined by the number of waveguide channels and the size of individual waveguides.

Our waveguide core is made of Parylene C, a high-refractive-index biocompatible polymer ($n = 1.639$), which is transparent throughout the visible range of the optical spectrum[42]. PDMS is used as the waveguide cladding due to its lower refractive index than that of Parylene C ($n = 1.4$). The material choice of Parylene C and PDMS provides a large index contrast ($\Delta n = 0.239$) among biocompatible polymers to confine an optical mode. A large index contrast improves mode confinement and results in a small bend loss. However, a large index contrastcan also exacerbate the scattering losses due to sidewall roughness, which can be alleviated by smoothing the waveguide sidewalls. A cross-sectional schematic of the waveguide structure on the wafer is shown in Fig. 1b.

### Embedded micromirrors enable vertical input/output coupling

A unique feature of our Parylene photonics implementation is the monolithic integration of embedded 45-degree micromirrors at the input and output ports, which enables broadband input/output coupling of light. The mirror topography is first precisely defined in a silicon mold (Fig. 1c) and then transferred to the flexible polymer device via deposition of Parylene C and PDMS onto the Si mold. These monolithically embedded micromirror structures are capable of 90-degree out-of-plane input/output light coupling (Fig. 1d).

As output ports, these micromirrors enable out-of-plane illumination normal to the surface of the implantable probe. Traditional optical waveguides and fibers operate in an end-firing configuration (Fig. 1e), in which light is emitted from the end facet. End-firing waveguides result in an in-plane beam profile that is oriented along the axis of the probe, causing a large portion of the probe surface area to be illuminated and limiting the number of nonoverlapping output ports that can be arranged on the surface. To enable a high spatial resolution illumination pattern along the probe shank, an out-of-plane scheme is preferred (Fig. 1f). Additionally, in the context of neural probes, where electrical recording sites are also patterned on the surface of the shank, an out-of-plane beam profile avoids direct illumination of recording electrode sites, reducing the severity of photoelectric artifacts[43]. A comparison of in-plane and out-of-plane illumination profiles is presented in Figure 1e, f. In-plane mirrors have also been used for side-firing waveguides in order to increase the optical probe spatial resolution[44], but out-of-plane mirrors are required to collocate stimulation with surface electrode arrays on the same probe shank.

Output ports may be lithographically defined in the desired arrangement along the probe shank to suit the purpose of the intended experiment. Although Parylene photonics can be broadly used in any biomedical application, here, we focus on device design in the context of neural stimulation using optogenetics. For example, output ports may be spaced along the length of the probe shank to stimulate different regions of tissue (e.g., different layers of cortex) or placed in a dense grid for interrogation of neural circuits in the same region.

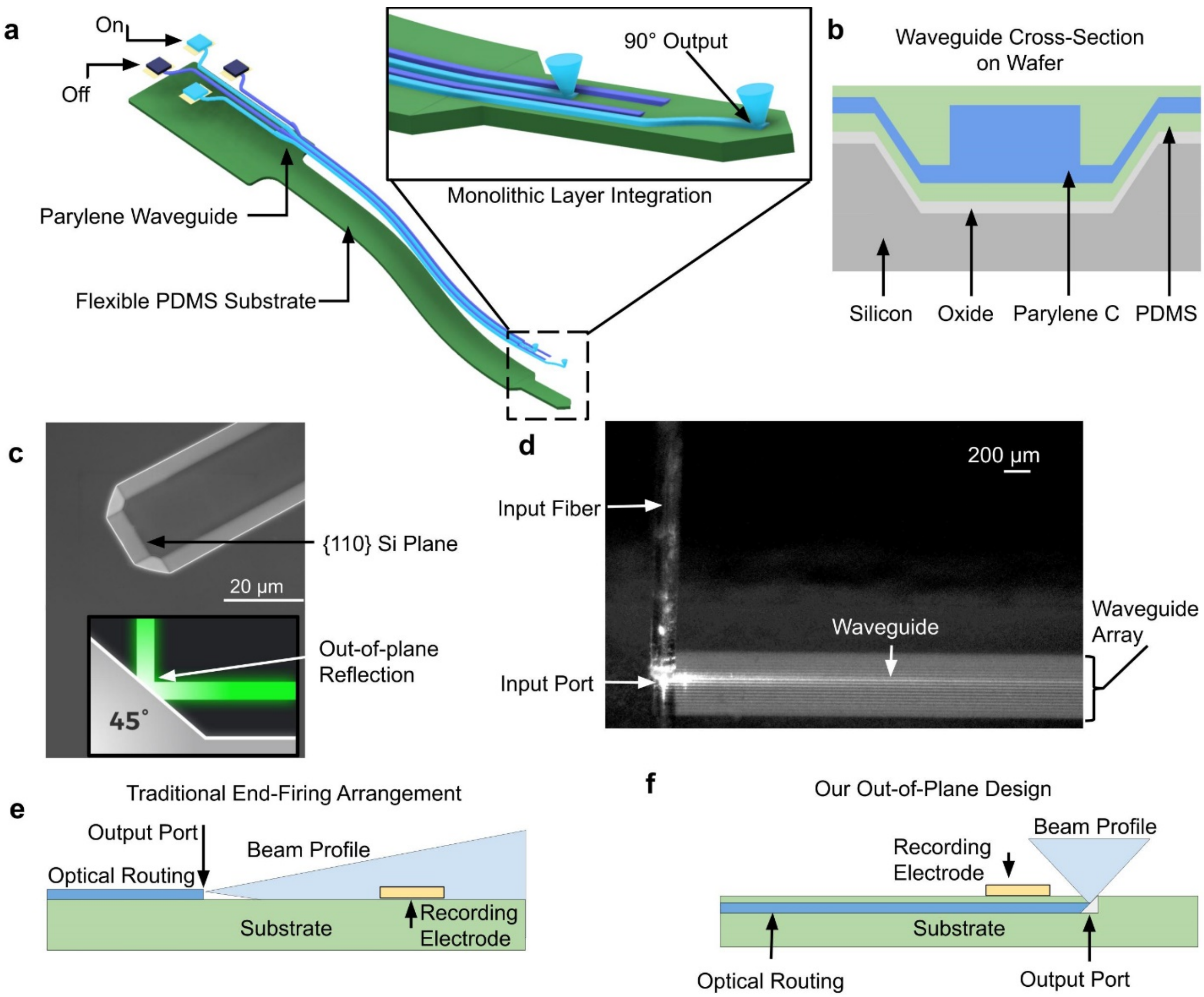

**Figure 1** a) Schematic diagram of the Parylene C/PDMS optical waveguide neural probe. The inset image shows the integrated device capable of out-of-plane light delivery. b) Schematic of the waveguide cross-section on the wafer. c) Scanning electron microscopy (SEM) micrograph of a Si trench etched via KOH with the {110} plane indicated, which acts as the mold for the micromirror port, as shown in the cross-sectional schematic diagram. d) Out-of-plane input coupling from an optical fiber into a Parylene C waveguide using the 45-degree micromirror at the input port. Adjacent waveguides appear bright due to the brightfield illumination and reflections from the bright beam spot of the input fiber. Only a single waveguide is excited, as seen by the bright line of outscattered light in the center of the array. Fiber alignment is performed using a precision micromanipulator (*PatchStar, Scientifica, UK*). e) Schematic of the traditional in-plane illumination from an end-firing waveguide, where waveguide illumination is along the probe shank, limiting the spatial resolution. f) Schematic of the out-of-plane illumination in our design, where the waveguide output is oriented perpendicular to the probe shank, allowing for higher spatial resolution.

**Packaging with optical fibers allows broadband operation**
Packaging of microfabricated optical waveguides with light sources is a uniquely stringent requirement for implantable applications. The device backend must be compact and robust to enable implantation. A unique feature of our integrated photonic waveguide platform is the embedded micromirror input ports that facilitate coupling of light from the surface.

Optical fibers can be aligned to the waveguide input facet using a 3D printed V-groove (Fig. 2a) and directly bonded to the waveguide array with optical epoxy, as shown in Fig. 2b (Materials and Methods). Due to the compact size of optical fibers (3.0 µm core diameter, 125 µm cladding diameter), many fibers may be bonded to the probe backend, allowing independent light coupling to multiple waveguides in the array. The chosen optical fiber (*S405-XP, Thorlabs Inc, USA*) operates as a single-mode fiber over wavelengths of 400 nm - 680 nm, covering the entire visible spectrum, relevant for most optical reporters and commonly used opsins for optogenetic stimulation. Packaging optical fibers at the backend of our implantable optical waveguide arrays has the advantage of enabling operation at different wavelengths using different external laser sources.

**Bonding laser diode chips enables compact backends**
It is highly desired that implantable photonic waveguide probes are realized such that the prohibitive tether connections to the backend are either eliminated or at least reduced in size to enable chronic experiments on free-roaming animals. Utilizing the micromirrors for input coupling, compact vertical-cavity surface-emitting laser (VCSEL) chips may be directly bonded to the input facet using a thin layer of anisotropic conductive film (ACF) (Material and Methods). The resulting probe is shown in Fig. 2c. The diode chips (*I0-0680M-0000-A002, Vixar Inc, USA)* emit at a wavelength of λ = 680 nm and are both compact (220 µm x 220 µm) and lightweight (0.5 mg). The ACF provides mechanical and electrical connections to the VCSELs without significantly attenuating the light. We measured an optical transmission of more than 67% through the ACF across the visible range of optical wavelengths. Direct integration of light sources precludes the need for an external fiber-coupled laser source. Thus, our optical probe requires only an external electrical connection to a pair of small wires, which can be much less cumbersome and restrictive than a brittle, delicate fiber connection.

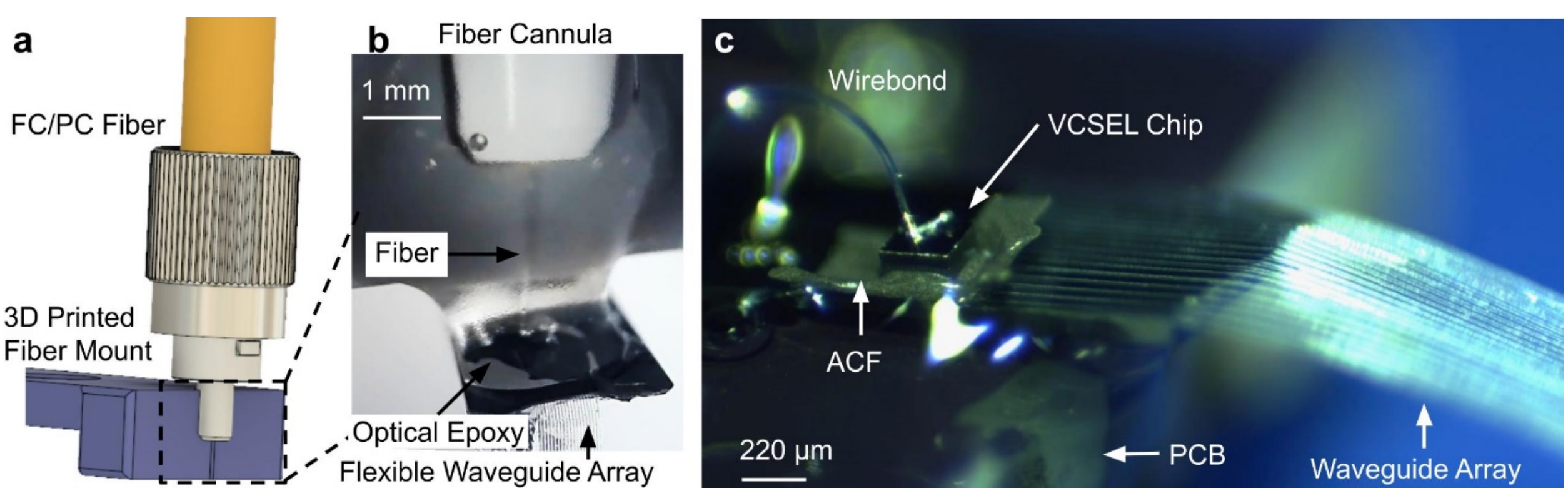

**Figure 2:** a) Schematic of a fiber-bonding 3D printed mount featuring a V-groove for fiber alignment. b) Fiber optic bonded to a flexible Parylene waveguide using optical epoxy. c) A vertical-cavity surface-emitting laser (VCSEL) chip bonded to the input port of a flexible Parylene waveguide using anisotropic conductive film (ACF).

## RESULTS
**Parylene photonic waveguides guide light with low propagation loss**
The Parylene photonics platform operates over a wide range of wavelengths, especially in the visible range that is relevant for optogenetic stimulation of neural activity. The input/output mirrors are broadband and enable coupling of light at different wavelengths (Fig. 3a). The propagation losses were measured at different wavelengths of interest for optogenetic stimulation including 450 nm (ChR2), 532 nm (Arch), and 633 nm and 680 nm (redshifted opsins, e.g., Chrimson)[48]. Table 2 lists the measured propagation losses in comparison with other waveguide technologies at different wavelengths. Measurement uncertainty is reported for the waveguide-to-waveguide measurement variation. These results show that the propagation losses of our Parylene C waveguides are comparable to those measured in silicon nitride waveguides used in neural probes (Table 2). Our previous work showed that the primary source of optical loss is the waveguide sidewall roughness as a result of etching the outline of the waveguide core[49]. Parylene photonic waveguides exhibit low losses across the entire visible range of the spectrum (450 – 680 nm), and the input/output coupling is broadband. Therefore, Parylene photonics can operate over a wide range of wavelengths.

**Table 2:** Comparison of Neural Probe Waveguide Material Platforms and Losses[1]

| Work | Material Platform | Wavelength λ (nm) | Propagation Loss α (dB/cm) | N (number of waveguides measured) |
|---|---|---|---|---|
| This Work | Parylene C | 450 | 6.1 ± 1.4 | 3 |
| Zorzos 2010[50] | SiN | 473 | 3.2 | |
| This Work | Parylene C | 532 | 4.9, ± 1.2 | 12 |
| This Work | Parylene C | 633 | 4.1 ± 0.8 | 14 |
| Kampasi 2016[51] | SiN | 635 | 5.0 | |
| This Work | Parylene C | 680 | 3.2 ± 1.7 | 5 |

**Flexible Parylene waveguides can guide light with low bend loss**

The Parylene photonics platform is designed as a flexible photonic architecture that can freely flex with the tissue to avoid exerting strain on the tissue (Fig. 3b). To operate reliably in vivo, the bends induced by tissue motion should not significantly impact the delivered optical power at the output port. However, bending of optical waveguides results in radiation of confined optical modes before light reaches the output facet. To characterize the overall bend loss, we experimentally measured the bend losses at different radii of curvature using a custom-designed jig (Fig. 3c). This jig has two parts. The first part is a 3D printed V-groove to hold the input fiber in place, and the second part is a precisely machined cylindrical rod to form the bend geometry. We used a series of different rods with different diameters to study the effect of bend loss at different bending radii.

[1] Measurement uncertainty is reported as ±1 standard deviation.

The experimental bend loss measurement results are presented in Fig. 3d, showing negligible bend losses for millimeter-scale bends, i.e., more than 95% of the maximum output intensity even at a bend radius of 1.5 mm. For bend radii smaller than 1 mm, the variance in the measurements is larger than that in the measurements at larger radii of curvature. This variance is caused by experimental challenges in wrapping the probe smoothly around rods at such small radii. We suspect that loops or creases formed in the probe shank during these tight bends reduce the measurement accuracy. Therefore, the presented flexible waveguides preserve their performance through millimeter-scale bends in the probe shank. Bends of this size (1.5 mm – 5 mm) are likely to occur during implantation and routing of the flexible shank in the body. Our bend loss measurement results suggest that the output optical power will be minimally affected by flexing in the tissue after implantation.

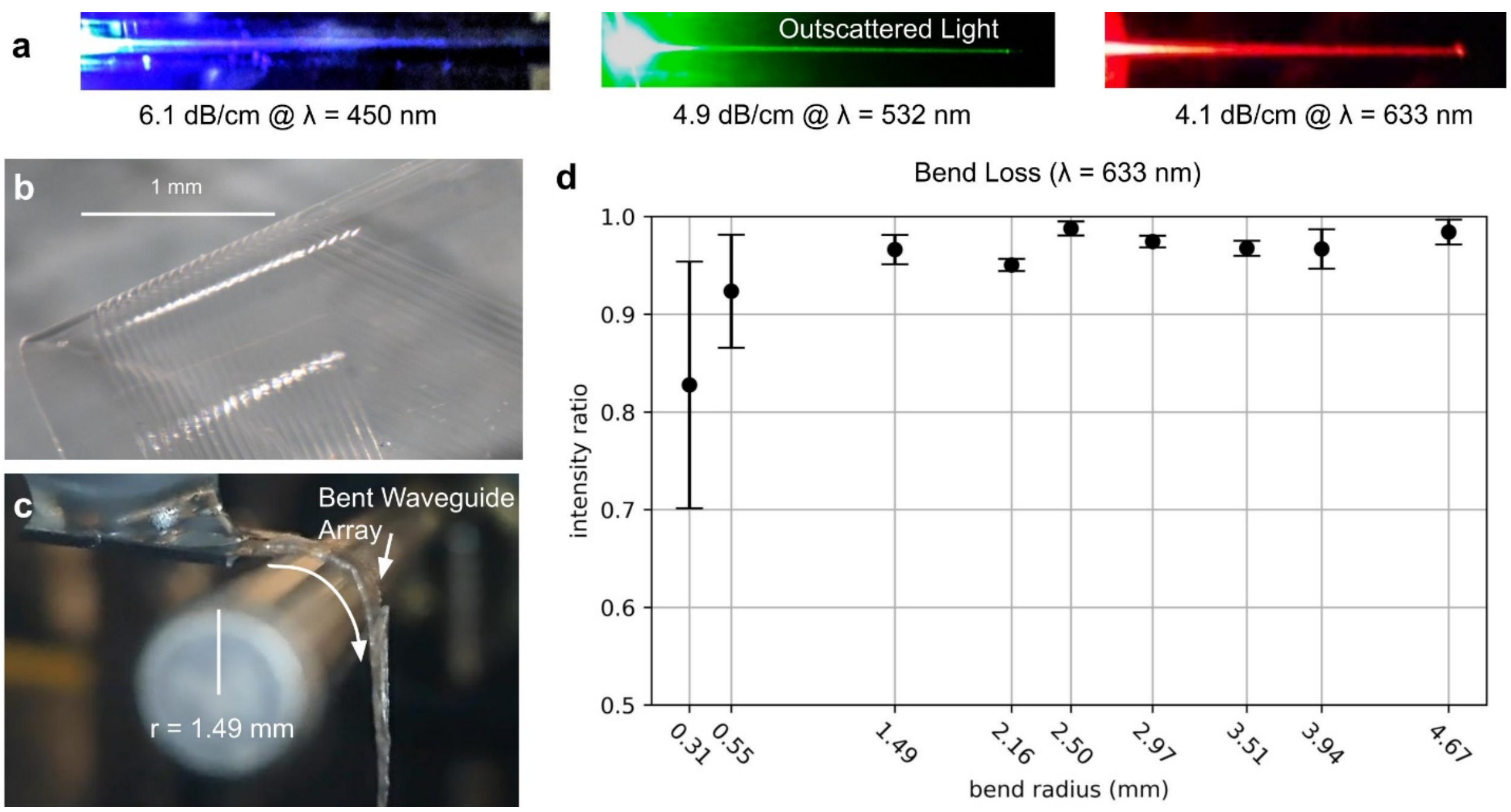

**Figure 3:** a) Outscattered light along the length of the waveguide is imaged at three different optical wavelengths in the visible range to show the trajectory of guided light. The waveguides were imaged from the side to avoid direct illumination from the output port, and the input power from the laser was increased so that the outscattered light along the waveguide path was clearly visible. b) Micrograph of the released Parylene C/PDMS waveguide array. c) The flexible waveguide array is bent over a custom jig to measure the bend loss (radius = 1.49 mm). d) Waveguide bend loss measurements of the relative output intensity (normalized to the output intensity from a straight waveguide) through a 90-degree bend of various radii. Low waveguide bend losses were demonstrated with a high intensity ratio (more than 95%) at millimeter-scale bends. The number of measurements for each datapoint was N = 4. The error bars denote the standard deviation.

**Embedded micromirrors enable out-of-plane illumination with localized beam profiles**

In addition to enabling broadband vertical input coupling, the 45-degree output micromirrors are capable of localized broadband illumination normal to the probe surface. To characterize the output beam profile, a charge-coupled device (CCD) camera was aligned to the output port of a Parylene waveguide in an array, and input coupling was adjusted to maximize the light intensity at the output port (Fig. 4a). Subsequently, to image the beam profile, a block of fluorescent tissue phantom (0.6% agar mixed with 10 ppm AlexaFluor 532) was aligned above the output port with a micromanipulator and imaged from the side (Fig. 4b). The resulting fluorescent emission profile is shown in Fig. 4c, with isointensity contours superimposed on the image to show the spatial decay of the light intensity.

The output beam profile reflected by the micromirror was quantitatively measured by imaging the output light intensity at multiple angles (Fig. 5d). Characterization of a 30 μm x 5 μm micromirror output port at λ = 532 nm shows a narrow beam profile (1/$e^2$ beam width is 13 degrees orthogonal to the surface of the probe (Fig. 5e, f). This localized illumination profile allows for multiple output ports to be independently spaced along the probe surface for targeted light delivery. The light intensity directly reflected from the output micromirror port was measured via a CCD camera to be > 36 dB more intense than the outscattered light from the waveguide.

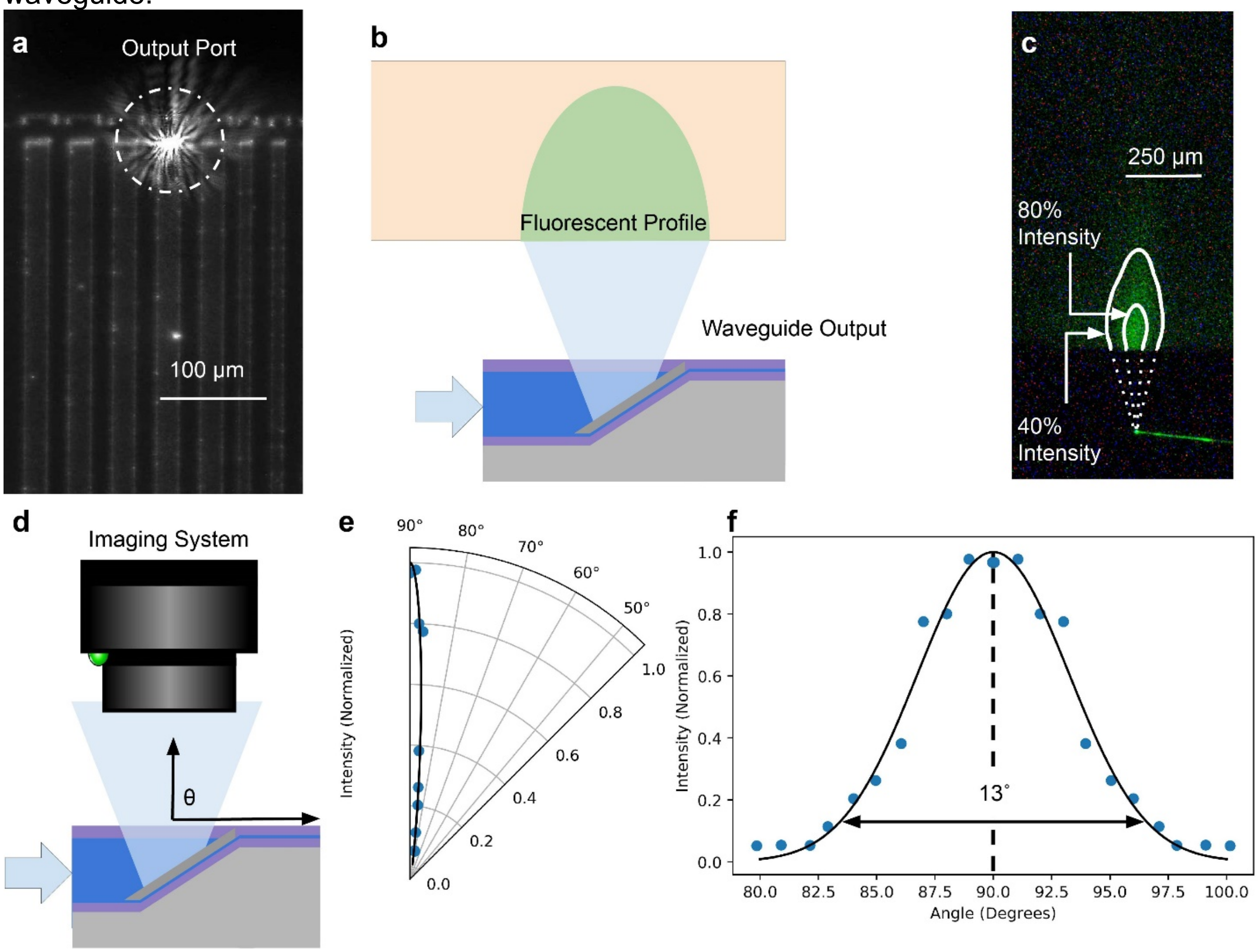

**Figure 4:** a) A brightfield image of the waveguide array (top view) featuring an illuminated output port. b) Schematic illustration of the fluorescent beam imaging experimental setup. c) Out-of-plane beam profile imaged in a fluorescent tissue phantom with labeled isointensity contours. d) Schematic of the beam profile characterization system. e) Radial beam profile, showing a peak intensity at 90 degrees, with rapid off-axis decay. f) Gaussian curve fit to the radial beam profile, showing a beam divergence of 13 degrees(1/$e^2$ beam width).

## Design of the cladding thickness enables integration of functional electrical layers

In the context of neural interfaces, both electrical recording and optical stimulation capabilities are desired to enable simultaneous electrophysiology recording and optogenetic stimulation experiments in the brain. Recording electrodes are usually formed via exposed metal sites connected by traces embedded in polymer insulation[45,46]. A conceptual schematic diagram of an additional planar layer of recording electrodes on a Parylene photonic waveguide is shown in Fig. 5a.

One concern of combining electrical and optical functionalities on the same platform is the interaction of the optical waveguide modes with electrical traces, which will decrease the delivered optical power due to absorption losses in the metal. Commonly used metals such as Au, Pt, Ti, and Al exhibit large absorption coefficients in the visible range of the optical spectrum[47]. In our device architecture, electrical traces can be routed along the length of the device, parallel to the optical waveguides. Therefore, any significant interaction between the guided optical mode and metal traces would cause significant attenuation of light after traversing the full probe length. This interaction can be minimized by routing the electrical traces through a separate layer, vertically spaced from the photonic layer by the PDMS cladding (Fig. 5b). To study the optical propagation loss due to the electrical traces, we performed rigorous finite difference eigenmode (FDE) simulations of the waveguide geometry (30 μm x 5 μm) with a thin sheet of metal (200 nm of Pt) situated over the cladding.

The Parylene photonic waveguide with dimensions of 30 μm x 5 μm is multimode in the visible range. Since each mode profile has a different spread outside and around the waveguide core, the modes experience different levels of attenuation from the presence of the metal sheet. However, when such a multimode waveguide is excited by an external light source (i.e., a laser), optical power is preferentially coupled to lower order modes due to the larger overlap between the typical Gaussian mode profile of the light source and the optical mode profiles of the lower order modes. Optical losses caused by interactions with the Pt layer were modeled for the lowest 30 modes and were found to be less than $3 \times 10^{-10}$ dB/cm for the fundamental mode and less than $5 \times 10^{-8}$ dB/cm for each of the remaining simulated modes (Fig. 5c). These results demonstrate that 1 μm of PDMS cladding is sufficiently thick to insulate the waveguide modes from interaction with additional metal layers.

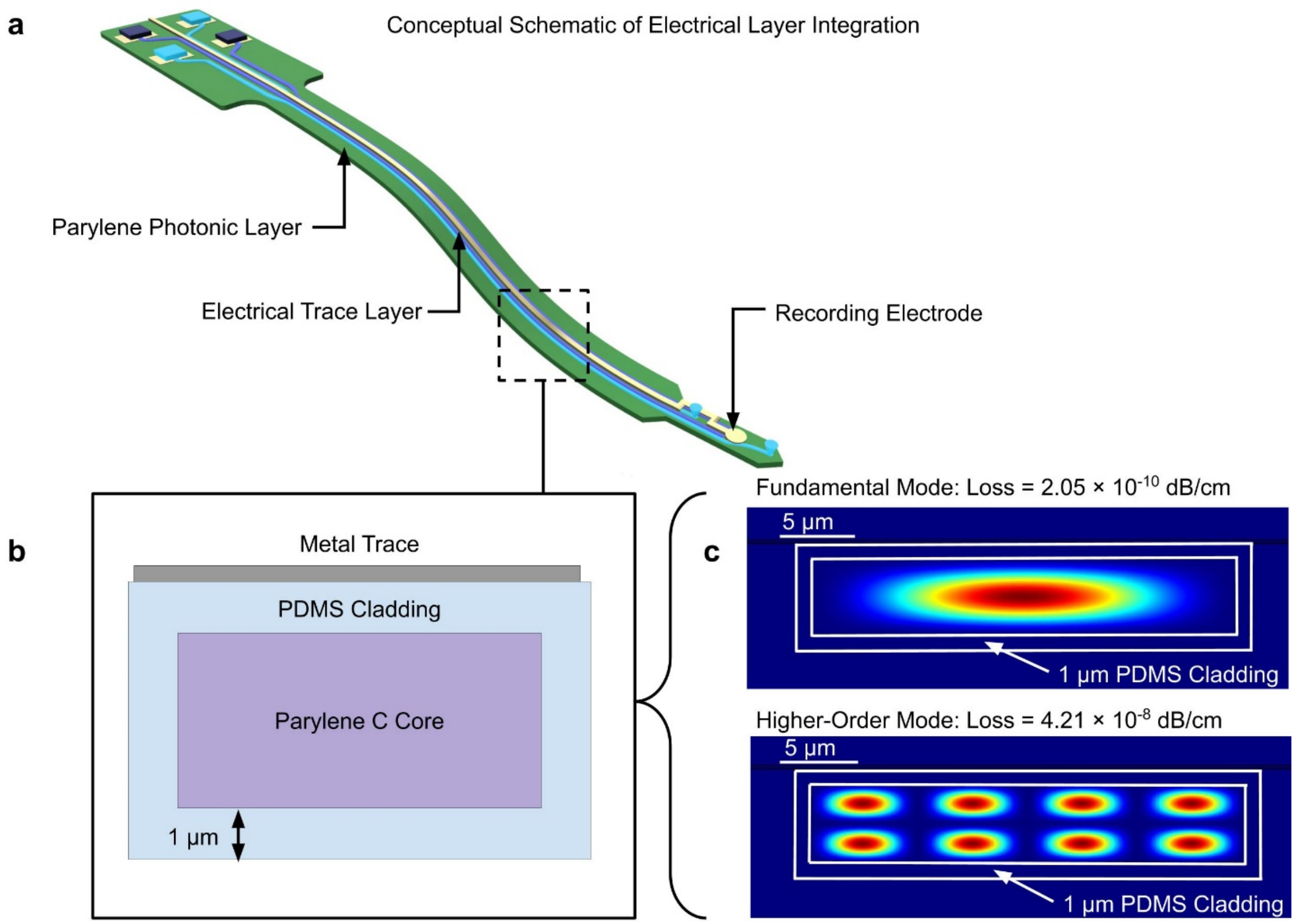

**Figure 5:** a) Schematic of a Parylene photonics neural probe with an additional layer of recording electrodes. b) Schematic cross-section of the device shank, with a metal trace over the waveguide core on top of the cladding. c) Mode simulations of losses due to the metal traces. Both the fundamental mode and higher-order modes show negligible losses due to interaction with the metal layer.

# DISCUSSION

The embedded micromirror input/output port is a unique feature of our platform. The micromirrors are broadband, unlike traditional out-of-plane illumination mechanisms such as grating couplers, which are highly wavelength dependent. Using the micromirrors, light at multiple wavelengths can be coupled to the waveguides. In the context of neural probes, different optogenetic wavelengths can be used to switch between stimulation and inhibition or to perform cell-type specific targeting. This Parylene photonics platform is the first to enable such high-resolution, broadband, out-of-plane light delivery in a fully compliant and biocompatible platform.

The waveguide output power can be controlled by changing the input optical power. Off-target illumination is minimal since the extinction between the output port light intensity and the outscattered background light from the probe shank near the output port is >36 dB. If the input power is very high, outscattered light along the probe shank can be further reduced by using additional optical shielding layers. For biological experiments, the input power must be carefully chosen to achieve an output power that is higher than the threshold of activation or detection of the optical agent of interest while also remaining lower than the threshold of photothermal damage to the tissue at the wavelength of operation.

Parylene photonics platform utilizes flexible polymer materials to reduce the tissue response after implantation. However, the overall device stiffness depends on the shape of the probe cross-section in addition to the material platform. For example, wide and thin probes are highly compliant when bent along the probe length but have higher stiffness along the width of the probe, resulting in a greater tissue response along the probe edges[52]. Therefore, a neural probe design with a compact footprint (i.e., thin and narrow), in addition to a soft material platform, is necessary to minimize damage to the tissue. When designing a specific implant using Parylene photonics, the number and size of optical channels must be chosen such that the overall width of the probe is minimized.

Here, we report relatively large waveguides (30 μm x 5 μm) for proof-of-principle demonstration. However, optical mode simulations show that the refractive index contrast between Parylene C and PDMS is sufficiently large to realize ultracompact optical waveguides that have well-confined modes even for small cross-sectional dimensions of 1 μm x 1 μm. At this small size, the mode exposure to the sidewall is increased, which necessitates process optimization to fabricate such devices with smooth sidewalls and reduce scattering losses. In a dense array configuration (1 μm gap), such waveguides exhibit negligible crosstalk of less than -30 dB over 5 cm length (Supplementary Material, Appendix I). These simulations suggest that under ideal conditions, the presented platform can be used to realize waveguide arrays even with an extremely dense pitch of 2 μm.

Another factor that limits the size and density of a multichannel device is the density of light source coupling at the device input. Here, we have demonstrated bonding of a single fiber to the waveguide input facet. Although the fiber core and cladding are small (125 μm diameter), serial bonding of individual fibers must take into account the prohibitive size of the fiber ferrule and its sleeve, which is typically 2.5 mm. Scaling the bonding process to many channels will require matching the waveguide spacing to the pitch of commercially available photonic chip coupler arrays, which are now available at channel pitches in the range of 20 μm to 127 μm to (*PLC Connections, USA)*.

In addition to coupling light from benchtop laser sources with a fiber tether, we leverage the versatility of the embedded micromirror input ports for direct out-of-plane coupling of light to the polymer waveguides from laser diode chips. The low weight of the VCSEL sources (0.5 mg/diode) is important in the context of chronic experiments on freely moving subjects, where the weight budget is typically 10% of the weight of the animal (2 – 3 g, headstage weight limit for mice[53]). This integrated laser diode platform may be directly modulated via electrical power supplies integrated into a headstage or used for tetherless experiments with the addition of a battery and radio frequency (RF) module. Due to the relatively large output facet (65 μm) of the bare chip VCSEL sources used here, a large input port is required to achieve efficient coupling into the waveguide. In future designs, the waveguide can be tapered to achieve high coupling efficiency while routing compact waveguides in the probe shank, or more compact laser diodes can be used.

The fabrication process outlined in this paper to realize Parylene photonics is compatible with commonly used microfabrication techniques. During the fabrication process, harsh chemicals such as hydrofluoric acid (HF) are employed as an efficient way to remove the oxide hardmask and sacrificial layers, necessitating careful rinsing in deionized water to avoid tissue contamination. Other hardmask and sacrificial release layers, such as germanium[54] (Ge), which can be removed using biosafe solvents such as 1% hydrogen peroxide, could be explored as alternatives. Our scalable microfabrication process enables monolithic integration of additional planar structures prior to release. Thus, using this platform, additional photonic layers can be stacked to increase device density, or electrical layers can be added to create a multimodal flexible device platform.

Although flexible devices are less damaging during chronic experiments, they are difficult to implant because they lack the structural rigidity needed to penetrate tissue. To address this need, implantation cannulas and bioresorbable stiffeners have been developed to temporarily increase the stiffness of neural probes for implantation. Bioresorbable materials such as polyethylene glycol (PEG), silk, callogen/gelatin, and PLGA have demonstrated tunable properties for both the stiffness and dissolution rate, which may be explored in combination with Parylene photonics[55]. The required insertion force can be reduced by shaping the bioresorbable stiffener into a sharp tip using controlled dip coating[56] or a molding process[57]. Additionally, external braces may be added to change the cantilever properties of the probe at the tissue interface to increase the buckling force of the probe[58]. These parameters require additional optimization of the Parylene photonics architecture, but compatibility with existing techniques allows future implementations to benefit from the rich literature on implantation techniques for flexible neural probes.

## CONCLUSION

Parylene photonics platform shows great promise for realizing chronic implantable biointerfaces due to its compliance, which can reduce the foreign body response in tissue. The out-of-plane, broadband input/output ports enabled by embedded micromirrors allow devices to create patterns of localized illumination beams normal to the surface for collocated integration with recording electrodes and enable direct packaging with light sources on the probe backend. This photonic device platform is broadband and offers unprecedented flexibility in choosing the desired wavelength of light for opsins and optical reporters. While in this paper, we discussed packaging and implantation considerations to show the feasibility of using this platform to realize optical neural probes, Parylene photonics can be used in a range of biomedical applications where a flexible, biocompatible optical device is desired.

## MATERIALS AND METHODS

### Fabrication process

**Preparing the silicon mold:** To implement Parylene C photonic waveguides and integrated micromirrors, fabrication was performed on 4-inch silicon wafers (n {100}) with 1 μm thermal oxide. The waveguide fabrication process flow is shown in Fig. 6 (with additional process details in Table 2). The thermal oxide layer serves as a hardmask for deep etching of silicon. Oxide was patterned using optical lithography and anisotropic reactive ion etching (RIE) (Step 1 in Table 3). Then, 45-degree Si sidewalls were formed using wet etching in potassium hydroxide (KOH) mixed with Triton X-100 surfactant[59] (Step 2 in Table 3) to reach the desired trench depth of 6 microns (Fig. 6a). The oxide hardmask was subsequently stripped via wet etching in 49% HF. Careful design of the mask orientation with features at 45 degrees to the (100) plane, indicated by the wafer main flat, is required to expose the (110) crystal plane and define the micromirror surface[59]. The patterned Si surface serves as a mold for subsequent polymer layers, defining the 3D shape of the micromirrors. Subsequently, 300 nm of conformal oxide was deposited on the patterned Si surface using a plasma-enhanced chemical vapor deposition (PECVD) process (Step 3 in Table 3) as a sacrificial layer to enable device release from the Si mold (Fig. 6b).

**Spin-coating PDMS substrate:** To form the substrate for the waveguide structure, a 1 μm PDMS layer was spin-coated on the silicon mold. Due to the high viscosity of PDMS, such a thin layer requires dilution with hexane prior to spin-coating. PDMS (*Sylgard 184, Dow Corning Corp, USA*) was diluted to 1:10 PDMS:hexane by volume, thoroughly mixed, and filtered through a 0.2 μm membrane filter to remove any particulates. The PDMS solution was degassed in vacuum (1 Torr) for 4 minutes to remove air bubbles. The solution was spin-coated for 60 s at 2000 rpm and then degassed again in 400 mTorr vacuum for 4 minutes. Finally, wafers were oven-baked for 45 minutes at 100 °C to cure the thin film and remove the solvent. To verify the thickness of PDMS to be 1 μm, the thin film was measured using a surface profilometer (*P-15 Stylus Profiler, KLA Tencor, USA*), wherein PDMS was mechanically removed from a portion of the wafer to measure the step height. The PDMS spin-coating process is not perfectly conformal and is affected by the waveguide trench topography. The spin-coating parameters and size of the trench must be optimized to achieve the desired thickness at the bottom of the trench. Multiple waveguides can be routed through a wide common trench.

**Metal micromirrors:** Due to the low surface energy of PDMS, photoresist cannot be directly spin-coated on its surface for lithography[60]. To overcome this issue, we developed and optimized a fabrication process in which a very thin (300 nm) layer of Parylene C film was deposited on PDMS to serve as an adhesion layer for the photoresist and enable optical lithography (Fig. 6c, Step 4 in Table 3). Direct chemical vapor deposition (CVD) of Parylene C on PDMS provides strong adhesion, making Parylene an ideal material[61]. Embedded metal micromirrors were patterned using a lift-off process consisting of lithography (AZ 4210) and evaporation of 5 nm Pt and 100 nm Al films (Steps 5,6 in Table 3). Pt serves as a strong adhesion layer to Parylene C[62], while Al is chosen as the mirror surface for its high reflectance across the visible spectrum[63]. Lift-off was performed via acetone soaking, followed by pulsed (5-10 s) sonication (Fig. 6d). The root mean square surface roughness of the Al micromirrors was measured via an optical profilometer (*Zygo NV7000, AMETEK, USA*) to be 49.3 nm with a standard deviation of 3.7 nm from mirror to mirror (N = 10).

**Waveguide core etching and smoothing:** In the next step, the waveguide core was realized in a subsequent layer of Parylene C. We fabricated multimode Parylene photonic waveguides with a core thickness of 3.5 μm. Parylene C was therefore deposited to a thickness of 3.5 μm using the CVD process described earlier. To define the outlines of individual waveguides, Parylene C was removed from the surrounding regions (Fig. 6e) using an anisotropic oxygen plasma etching process. We used a 40 nm sputtered chromium (Cr) hard mask (Step 7 in Table 3) to achieve a high selectivity for etching Parylene C. The waveguide patterns were aligned to the mirrors using a contact lithography process (AZ 5214E), and the hard mask was patterned by wet etching of Cr *(Cr 1020 Etchant, Microchem GmbH, DE*).

The patterns were then transferred to Parylene C via oxygen plasma RIE (Step 8 in Table 3). PDMS acts as an etch stop layer since this polymer is not effectively etched by oxygen plasma alone[64]. After Parylene C was etched (verified via reflectometry), the Cr hardmask was stripped with Cr etchant. The etched sidewall roughness results in optical scattering and significant propagation loss, thus rendering the optical waveguide impractical for efficient guiding of light. To address this issue, we used our previously reported technique of depositing an additional 1.3 μm conformal layer of Parylene C over the etched sidewalls to reduce the sidewall roughness and the associated propagation loss[49]. We use this technique here to smoothen the etched sidewalls and reduce the propagation loss of the implantable waveguides. The three sequential Parylene C layers, i.e., the thin layer on the PDMS substrate, the waveguide layer, and the conformal coating on the top, form the waveguide core with a total thickness of approximately 5 μm (Fig. 1b).

**Device release:** After the upper cladding of 1 µm PDMS was spin-coated (Fig. 6f), a 1 µm aluminum (Al) hardmask was sputtered to define the outline of the entire waveguide array (Step 9 in Table 3). The Al hardmask was lithographically patterned (AZ 4210) and wet etched (*Al Etchant Type-A, Microchem GmbH, DE*). PDMS cladding was etched, and arrays were singulated using RIE (Step 10 in Table 3). Finally, the Al hardmask was stripped. To release the devices, the silicon substrate was first thinned down to 100 µm using backside etching in $SF_6$ (Step 11 in Table 3), and then, the thinned Si wafer was completely etched in a subsequent etching step in $XeF_2$ (Xactix e2). The sacrificial oxide layer serves to protect the backside of the waveguide array. Once Si was removed, the sacrificial layer was stripped in buffered hydrofluoric acid (BHF), resulting in a released, flexible waveguide array (Fig. 6g). The devices were thoroughly rinsed in deionized water after release to avoid contamination of biological tissues by the process chemicals.

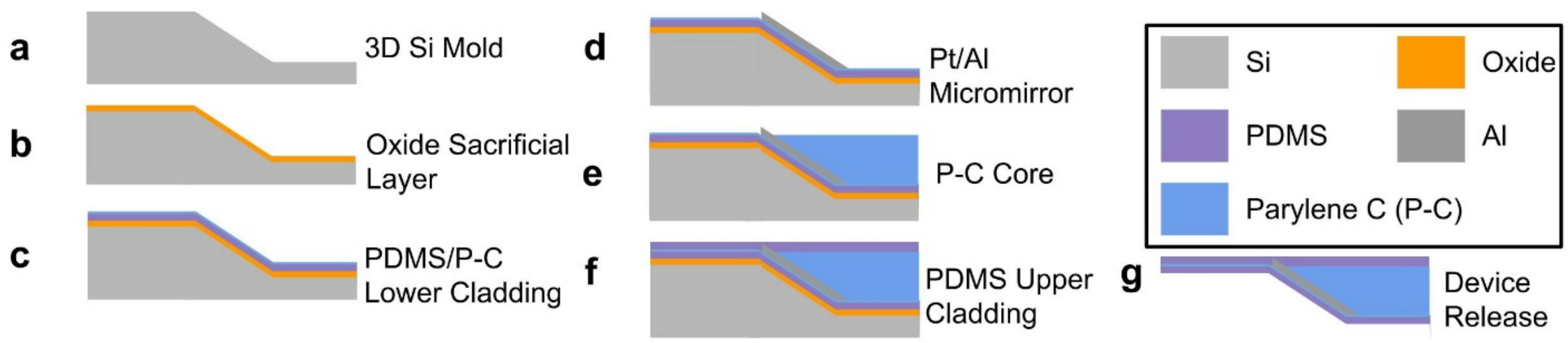

**Figure 6:** (a - g) Main steps to realize Parylene photonic waveguides.

**Table 3:** Fabrication Process Parameters

| Process Step | Tool | Parameters | Rate |
|---|---|---|---|
| 1) Thermal Oxide Etch | PlasmaTherm 790 RIE | Gas: 22.5 SCCM $CHF_3$<br>Gas: 16 SCCM $O_2$<br>Pressure: 100 mTorr<br>Power: 200 W | 55 nm/min |
| 2) Anisotropic Si Etch | Wet Bench | Concentration: 2 M KOH<br>Concentration: 60 ppm Triton X-100 Surfactant<br>Temperature: 90 °C<br>Agitation: 210 rpm stirring | 280 nm/min |
| 3) PECVD Oxide Deposition | Trion Orion II PECVD | Temperature: 375 °C<br>Pressure: 900 mTorr<br>Gas: 75 SCCM $N_2O$<br>Gas: 70 SCCM $SiH_4$ | 60 nm/min |
| 4) Parylene Deposition | SCS Labcoter-2 | Furnace Temperature: 690 °C<br>Chamber Gauge Temperature: 135 °C<br>Vaporizer Temperature: 175 °C<br>Pressure: 35 mTorr | 1 g -> 300 nm<br>2 g -> 1.3 µm<br>6.9 g -> 3.5 µm |
| 5) Pt Evaporation | Kurt J. Lesker PVD 75 Electron Beam Evaporator | Pressure: $3 \times 10^{-7}$ Torr | 3 Å/s |
| 6) Au Evaporation | Kurt J. Lesker PVD 75 Electron Beam Evaporator | Pressure: $3x10^{-7}$ Torr pressure | 5 Å/s |
| 7) Cr Sputtering | CVC Connexion Sputtering System | Pressure: 7 mTorr<br>Power: 50 W RF<br>Gas: 50 SCCM Ar | 10 nm/min |
| 8) Parylene Etching | Trion Phantom II RIE | Gas: 14.0 SCCM $O_2$<br>Pressure: 50 mTorr<br>Power: 50 W RF | 250 nm/min |
| 9) Al Sputtering | Perkin Elmer 8L Sputtering System | Power: 100 W RF<br>Pressure: 5 mTorr | 30 nm/min |
| 10) PDMS Etching | Trion Phantom II RIE | Pressure: 75 mTorr<br>Gas: 30 SCCM $CF_4$<br>Gas: 10 SSCM $O_2$<br>Power: 200 W RF | 200 nm/min |
| 11) Si Etching | Trion Phantom II RIE | Pressure: 25 mTorr<br>Power: 100 W<br>Gas: 30 SCCM $SF_6$ | 2.3 μm/min |

## Experimental Methods

**Optical fiber bonding:** First, a drop of optical quality epoxy (*EPO-TEK 301, Epoxy Technology Inc, USA*) was placed on the waveguide array backend. An optical fiber was fixed to a custom-designed 3D printed fixture with a V-groove and aligned to the input micromirror using a precision micromanipulator (*PatchStar, Scientifica Inc, UK*). After maximizing the input coupling efficiency, the epoxy was thermally cured (60 °C, 2 h) to provide a stable mechanical connection between the fiber and the waveguide array.

**VCSEL light source bonding:** ACF film (*CP34531-18AK, Dexerials Corporation, Tokyo, Japan*) was placed over the input micromirror. The VCSEL chip was then aligned to the input facet using a commercial flip-chip bonding tool (*M9A, BE Semiconductor Industries N.V. (Besi), The Netherlands*). Once aligned, the ACF was cured (120 °C, 15 minutes) to fix the VCSEL chip in place. The p-contact and n-contact of the diode were electrically connected to an external PCB using an Al wirebond (*Model 7476D, West Bond, Anaheim, CA*).

**Characterization of propagation loss:** To characterize the device performance, a single-mode fiber (*S405-XP, Thorlabs Inc, USA)* or a laser diode (*PL 450B, OSRAM GmbH, Germany*) was aligned to the input facet of the waveguide while imaging the output port onto a CCD camera. The input coupling was optimized by adjusting the position of the light source using a precision micromanipulator *(PatchStar, Scientifica Inc*) to maximize the light emission from the output port. Propagation losses in the waveguide were characterized by measuring the decay in the intensity of outscattered light along the length of the optical waveguide[49]. Traditionally, the modified cutback method is used to measure waveguide loss, which involves recording the output light intensity from multiple waveguides of different lengths and interpolating the loss as a function of length. However, our method based on extracting the propagation loss from the outscattered light allows for measurements of individual waveguide performance in one shot and thus eliminates the errors inherent to the cutback method due to waveguide-to-waveguide variations or changes in the input coupling efficiency from one waveguide to another.

**Characterization of bend loss:** A single Parylene photonics probe was sequentially wrapped around rods of different radii (0.3 mm – 5 mm) to measure the bend loss. In all cases, the bend angle was 90 degrees. The transmitted output power was measured by imaging the output port onto a CCD camera. To minimize errors due to variations in input coupling conditions, we optimized and maximized the coupling from the input optical fiber by measuring the transmission of the straight waveguide and then permanently affixed the fiber to the input facet using an epoxy to maintain consistent input coupling. We repeated the experiments four times, each time sweeping the range of bend radii, to ensure the repeatability of the results and characterize the measurement error at different radii.

**Characterization of the micromirror output beam profile:** The optical properties of the waveguide platform were characterized to demonstrate broadband, localized, out-of-plane illumination capabilities. A single-mode optical fiber *(S405 XP, Thorlabs Inc, USA)* was aligned vertically to the input mirror for input coupling to the waveguide.

The input coupling efficiency was optimized by adjusting the fiber position using a precision micromanipulator (*PatchStar, Scientifica Inc.*) while monitoring the peak intensity from the output port in the CCD camera (*EO-5012M, Edmund Optics, USA*).

Output port beam profiles were imaged by rotating the imaging platform in fixed increments about the axis of the waveguide and imaging the output port light intensity. To increase the dynamic range of the system, the imaging exposure time was scaled for each sample to avoid individual pixel saturation, and the pixel intensity values were then scaled by the exposure time.

**Agar phantom preparation:** Powdered bacteriological agar (*A5306, Sigma-Aldrich, USA*) was mixed by weight with 200 mL of deionized water in a 500 mL beaker and stirred at 600 rpm while heating on a hotplate. The top of the beaker was covered with a quartz watch glass to prevent evaporation. Once the solution was boiling, the hotplate was switched off, and the temperature was monitored via a thermocouple until dropping to 60 °C. For fluorescent samples, 10 ppm AlexaFluor 532 was added to the solution and mixed for an additional 5 minutes. The solution was then poured into a mold and cooled in a refrigerator at 6 °C for 30 minutes before use.

## Simulation method

Optical simulations were performed using commercial FEM software (*Lumerical MODE Solutions 2018b, Lumerical Inc, USA)*. The waveguide mode profiles were solved in 2 dimensions (2D) by taking a cross-section of the device geometry along the axis of propagation (z-axis). The finite difference eigenmode method was used to solve Maxwell's equations in 2D to find the electric and magnetic field profiles for each mode, as well as the complex propagation constant, β. The mode profiles were visualized by plotting the normalized electric field intensity, $|E|^2$. A perfectly matched layer (PML) boundary condition was used around the simulation domain. In all simulations, a background index of 1.0 was used. Simulations were performed at a wavelength of $\lambda$ = 450 nm.

Parylene waveguides were simulated with dimensions of 1 μm x 1 μm and 30 μm x 5 μm.

To simulate the waveguide structure, Parylene C is defined as a dielectric with refractive index of n = 1.639, and PDMS is defined as having a refractive index of n = 1.4. The platinum optical properties were taken from E. Palik[65].

# Acknowledgments

This material is based upon work supported in part by the National Science Foundation under Grant No. 1926804 and also by the National Institute of Neurological Disorders and Stroke of the National Institutes of Health under Award Number 1RF1NS113303. The authors acknowledge support by the Carnegie Mellon University Ben Cook Presidential Graduate Fellowship (J.W.R.), the Carnegie Mellon University Richard King Mellon Foundation Presidential Fellowship in the Life Sciences (J.W.R.), the Axel Berny Presidential Graduate Fellowship (J.W.R), Philip and Marsha Dowd (J.W.R), the National GEM Consortium (M.L.). The authors acknowledge the support of the Carnegie Mellon Nanotechnology facility.

# Conflict of Interest

The authors declare no conflicts of interest.

# Contributions

JR designed the devices, performed microfabrication, characterization, modeling, simulation, and prepared manuscript and the figures. ML performed characterization, packaging, and prepared the figures. MC served as the principal investigator on the project, designed and supervised the research, and prepared the manuscript. All authors reviewed the manuscript.

## Appendix I: Crosstalk Simulation

Parylene C and PDMS are excellent candidate materials for a biophotonic platform due to their biocompatibility, flexibility, transparency in the visible range, and ease of processing using traditional microfabrication techniques. Additionally, Parylene C has a high refractive index among polymer materials (n = 1.639), whereas PDMS has a low refractive index (n = 1.4), allowing for devices with high index contrast. Compared to dielectric materials such as silicon nitride (n = 2.05 at λ = 450 nm)[66], however, Parylene C provides much less index contrast. A simulation is performed to show that the index contrast in Parylene C/PDMS is sufficient for a high-density photonic platform. A concern of densely integrated waveguide arrays is the crosstalk between adjacent waveguide channels due to evanescent field coupling between the channels. As waveguides become smaller, the optical modes become less confined, and more optical power propagates in the cladding. As a result, this optical power can be coupled into adjacent waveguide channels.

The simulation is performed using a bidirectional eigenmode expansion (EME) solver (*Lumerical MODE Solutions 2018b, Lumerical Inc., USA)*. The simulated waveguide has a cross-section of 1 µm x 1 µm (Fig. S1a). Two parallel waveguides with 1 µm spacing are simulated over a propagation length of 5 cm (Fig. S1b). The input optical power in the "input waveguide" is normalized to 1, and the proportion of optical power in the adjacent "probe waveguide" is measured. Over the 5 cm propagation length, increasing power is observed in the probe waveguide due to evanescent field coupling (Fig. S1c,d). Over a 5 cm length, a total crosstalk of -32.86 dB is observed.

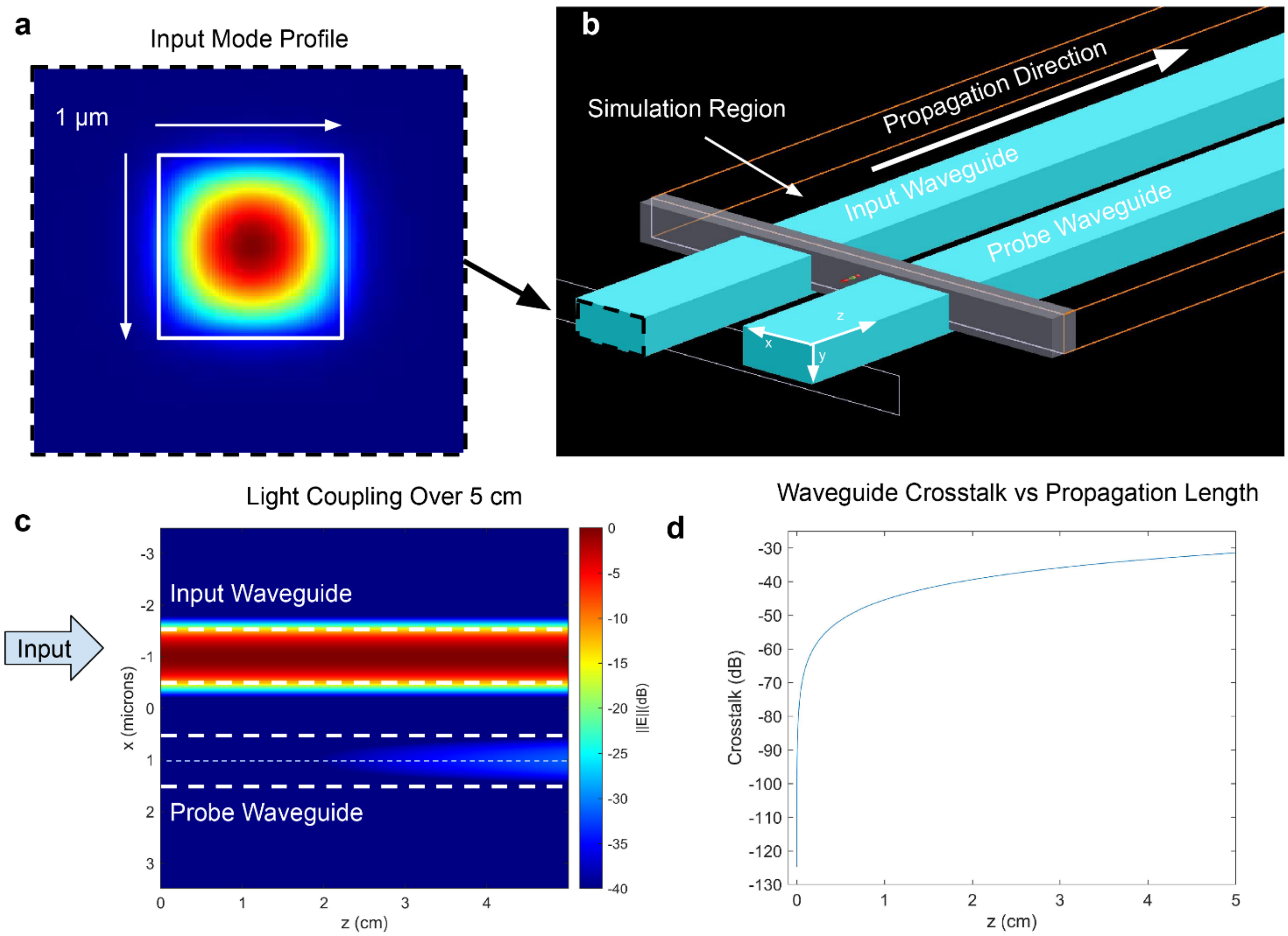

**Figure S1:** a) The optical mode profile of a 1 µm x 1 µm waveguide showing the fundamental mode at λ = 450 nm. b) Schematic of two parallel waveguides for crosstalk analysis. The input waveguide (left) is powered while the probe waveguide (right) is monitored over a propagation length of 5 cm. c) Electric field profile of the parallel waveguide structure over a 5 cm propagation distance. Light coupling into the probe waveguide is observed. d) Plot of crosstalk in the probe waveguide vs propagation length in decibels (dB). At 5 cm, a crosstalk of -32.86 dB is observed.